\documentclass[reprint,amsmath,amssymb,aps]{revtex4-2}

\usepackage{graphicx}
\usepackage{dcolumn}
\usepackage{bm}
\usepackage{chemformula}
\usepackage[T1]{fontenc}
\usepackage{upgreek}
\usepackage[mathlines]{lineno}

\begin{document}

\preprint{APS/123-QED}

\title{Frequency Dependent Dewetting of Thin Liquid Films Using External ac Electric Field}

\author{Bidisha Bhatt}
\author{Soumik Mukhopadhyay}
\author{Krishnacharya Khare}
\email{kcharya@iitk.ac.in}
\affiliation{Department of Physics, Indian Institute of Technology Kanpur, Kanpur, India-208016}

\begin{abstract}
Stability of thin liquid films on a surface can be controlled using an external stimuli, such as electric field, temperature or light by manipulating the total excess free energy of the system. It has been previously shown that thin lubricating films on slippery surfaces can be destabilized via spinodal mechanism using external electric field, which return to the original stable configuration upon switching off the electric field. However, the role of the frequency of the applied electric field is not clear, which is the main topic of study in this report. When an ac electric field of fixed voltage and varying frequency is applied across thin lubricating films of slippery surfaces, different dewetting behavior is observed. Characteristic length and time scales of dewetting depend strongly on the frequency of the applied voltage, which is primarily due to the change in the dielectric behavior of the lubricating fluid. In addition, the interplay of various time scales involved in the dewetting process also depend on the frequency. 
\end{abstract}

\keywords{Suggested keywords}

\maketitle

\section{Introduction}
The evolution of the thin film morphology and its control during dewetting attracted growing research interest for fundamental and technological applications. \cite{gentili2012applications, Alizadeh2018Thinfilms, telford2017functional, redon1991dynamics} Thin films are widely used in biomedical to make the easy flow of complex liquids, dielectric layers for electrical applications and optical applications, and many more. \cite{walheim1999nanophase} The stability of the thin films is crucial for these applications; the rupture of the film leads to the failure of the device. Dewetting is the dynamic process of retracting the liquid film from a surface due to intermolecular forces or external perturbations such as electric field, temperature, and many more. \cite{peschka2019signatures, mitov1998convection, warner2002dewetting, alvarez2008surface, sterman2017rayleigh, kataoka1999patterning, schaeffer2000electrically, surenjav2009manipulation, usgaonkar2021achieving, seemann2001dewetting, seemann2005dynamics} Dewetting is often observed in various industrial and natural surfaces, such as coating and printing processes, lubrication, and droplet coalescence. \cite{schwartz2001dewetting, junisu2022film} The fundamental study of the dewetting dynamics and its final morphology is necessary to use dewetting for potential applications.

Thin films are also interesting because they provide frictionless motion to liquids for various applications. \cite{lafuma2011slippery, wong2011bioinspired} The drop manipulation over thin films gains attention because of its very low ($<2^{\circ}$) contact angle hysteresis, and these surfaces are known as slippery surfaces. However, the liquid films are sandwiched between the drop and substrate, stable for the hydrophobic substrate but dewet on hydrophilic surfaces. \cite{daniel2017oleoplaning, bhatt2022dewetting} The total excess free energy of the liquid film underneath the aqueous drop depends on the intermolecular interactions between different mediums. \cite{sharma1993relationship, reiter1999thin} The polymers used to prepare are dielectric, so the excess free energy of the film can be tuned by using the external electric field. \cite{schaffer2001electrohydrodynamic, verma2005electric} Also, aqueous drop deposited on thin films shows the change in contact angle with response to the applied electric field, known as electrowetting. \cite{chen2014electrowetting, mugele2005electrowetting} 

Andelman et al. published an article on electrowetting for an ac-applied field that captured the change in contact angle of the aqueous drop with a range of frequencies and voltages. \cite{klarman2011model} For electrowetting on slippery surfaces, the contact angle depends on the strength ($V^2$) of the external electric field and follows the Young-Lippmann (Y-L) equation. \cite{barman2017electrowetting, armstrong2020evaporation} However, before contact angle saturation, the contact angle follows the Y-L equation, but a recent study shows that on slippery surfaces while applying an electric field, the underneath film breaks into small dewetted droplets. \cite{staicu2006electrowetting, bhatt2022dewetting} The dewetting process is reversible by switching On and OFF the voltage, but it takes longer to rewet the film after switching OFF the voltage. \cite{edwards2016not, edwards2020viscous, bhatt2022dewetting} By applying the electric field across the dielectric liquid film, the amplitude of the capillary waves present on the surface gets amplified and follows the linear stability analysis (LSA). In LSA, the contribution of the frequency of the applied field and its response, i.e., dielectric relaxation of the polymer, is not included. \cite{verma2005electric, bhatt2023electric} The system consists of the two conducting layers, an aqueous drop and substrate, and the dielectric film is sandwiched between them; this whole arrangement is like the parallel plate capacitor filled with dielectric. After applying an electric field, the dielectric film gets polarized with the time constant (charge relaxation time) and shows different relaxation dynamics such as $\alpha$-, $\beta$- or $\gamma$-relaxations, depending on the applied frequency range. \cite{schroeder2002segmental, raju2003dielectrics} 

The dielectric response of the different materials in the system also modified the excess free energy, which is not included in the LSA. Some studies show that the frequency response of the system follows the non-linear and Floquet theories.\cite{gambhire2012role, roberts2009ac, nesic2015fully} Now, due to this, the charge stored in the dielectric layer and the time constant changes the total energy of the system. So, only using amplitude in the Young-Lippmann equation will not give the correct electric response of the drop.

Herein, We investigate the effect of applied ac frequency on the dewetting morphology and the dynamics. We have found that, with the applied frequency, the electric properties, viz., dielectric relaxation of the material, are responsible for the change in the fastest-growing wavelength and growth of the capillary waves. Change in dielectric properties is responsible for the change in the excess free energy of the film underneath the aqueous drop. We have modified the LSA for the frequency response of the applied ac field and found that the modified theory matches the experimental observations. Understanding dewetting dynamics with applied electric field strength as well as frequency will be helpful in performing drop manipulation over these surfaces. And one can choose the voltage and frequency range such that the underneath film is stable for applied field and easy drop motion on such surfaces.

\section{Experimental Section}
\subsection{Materials}
Silicon Wafers having 1 $\upmu$m thick silicon oxide layer (p-type, $<$100$>$, resistivity 0.001-0.005 ohm-cm) were purchased from University Wafers Inc.. The breakdown voltage for 1 $\upmu$m SiO$_{2}$ is 300 V$\upmu$m$^{-1}$, is used to prevent the circuit from an electrical breakdown. Wafers were diced into 2 cm$\times$2 cm and thoroughly cleaned with ethanol, acetone, and toluene in an ultrasonic bath followed by oxygen plasma (Harrick Plasma) for 5 min. To satisfy the stability condition of the slippery surfaces, the surface energy of the silicon oxide surfaces was modified by grafting a self-assembled monolayer of octadecyltrichlorosilane (OTS, Sigma-Aldrich) molecules. The contact angle of water on freshly prepared OTS grafted surface is 106$^{\circ}$($\pm2^{\circ}$). The base of sylgard 184$^{\mathrm{TM}}$ (Polydimethylsiloxane (PDMS), Dow Corning, viscosity 5000 cSt, surface tension 21.2 mNm$^{-1}$) was used as lubricating fluid. For fluorescence imaging, oil miscible dye (Nile red, Sigma-Aldrich) was added with lubricating fluid (PDMS). An aqueous solution of 80$\%$ Glycerol (Fisher Scientific) and 20$\%$ DI water (with 0.1 M NaCl) was used as a test liquid. NaCl was added to increase the conductivity of the test liquid to make it responsive to an applied electric field. Above composition of glycerol and water is hygroscopically stable at our experimental parameters. To apply an electric field across PDMS film, copper wire (diameter 70 $\upmu$m) was connected with the silicon substrate by using a silver paste (Fisher Scientific), and a platinum wire (diameter 250 $\upmu$m) was inserted inside aqueous drop. All materials were used as it is without any further purification.

\subsection{Preparation of slippery surface and dewetting experiments} 
Cleaned diced silicon substrates were immersed in a 0.2 V/V$\%$ of OTS solution in toluene for 20 min. After taking out from the solution, substrates were ultrasonicated for 5 min in toluene to remove non--grafted OTS molecules and then heated for 30 min. at 90$^\circ$C. Copper wire was pasted at the edge of the sample by using conducting silver paste (Sigma-Aldrich) and leaving it to air dry for 1 h. To make a thin film of lubricating fluid for dewetting experiments, PDMS was diluted with n-heptane (having Nile red dye with 0.0015 W/V$\%$) with a 4 W/V$\%$ ratio. The diluted solution was poured on OTS grafted substrate and spin coated for 100 s with 2000 RPM and 10 s acceleration. The thickness of the prepared film was measured using an optical profiler (F-20, KLA USA) is $500\;(\;\pm\;10)$ nm.

\subsection{Dewetting experiments} 
An aqueous drop of 0.5 ml volume was deposited on a 500 ($\pm$10) nm thin PDMS film, and the system was chosen such that the PDMS film was stable underneath the aqueous drop. To observe the effect of applied frequency on dewetting dynamics, ac electric field was generated from a function generator (SG1610C, Aplab India). The signal amplitude was amplified using the high--voltage amplifier (T-50, Elbatech Italy). The frequency and amplitude of the applied voltage were measured using a digital oscilloscope (GDS-1062, Gwinstek India). ac signal was applied between the aqueous drop and the bottom silicon substrate. The dewetting dynamics were observed using the fluorescence optical microscope (BX-51, Olympus Japan) having a color CMOS camera (10 fps, $1024\;\mathrm{pixel}\;\times\;798\;\mathrm{pixel}$). Image analysis was done by using the open--source software, Gwyddion, and ImageJ. To relate the microscopy experiments with the polymer relaxation dynamics with applied field, dielectric spectroscopy (Precision Premier II) measurements were done. The dielectric constant and loss measurements with applied frequency were done using the parallel plate arrangement of two copper plates dipped inside the solution (PDMS or aqueous). Dielectric measurement data were analyzed using licensed ``Vision" software.

\section{Results and discussion}
\subsection{Effect of the frequency of applied ac voltage}
\begin{figure*}[ht!]
	\centering
		\includegraphics[width=1.0\textwidth]{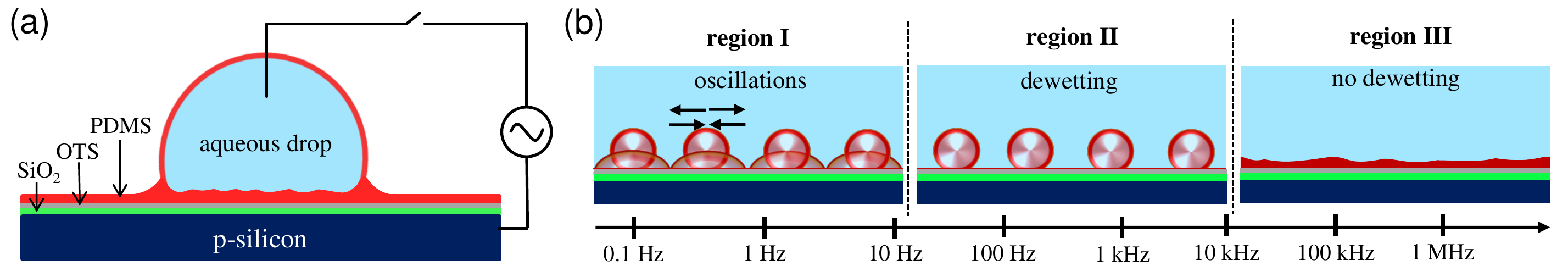}
	\caption{Schematics showing (a) the experimental setup and (b) dewetting morphology divided into three regions based on the frequency of the applied ac voltage. In region I ($f <$ 10 Hz, films and dewetted droplets oscillate as they can respond to the low frequency. In region II (10 Hz $< f <$ 10 kHz), stable dewetting pattern is observed, whereas in region III ($f >$ 10 kHz), no dewetting is observed.} 
	\label{Fig:Figure 1}
\end{figure*} 
Figure \ref{Fig:Figure 1}(a) shows the schematics of the experimental setup to study the frequency dependent stability (dewetting) of thin lubricating films underneath aqueous drops on slippery surfaces. In the absence of external electric field, stability of the liquid films depend only the apolar (Lifshitz-van der Waals) and the polar (acid-base) intermolecular interactions. It can be defined in terms of the total excess free energy per unit area of the films underneath aqueous drops. For thin PDMS films on OTS grafted SiO$_2$ substrate, the Lifshitz-van der Waals interaction is solely responsible for the stability of the films; the polar contribution is negligible because of the apolar nature of the films. When the external electric field is applied across the PDMS films, an extra contribution in the excess free energy (per unit area) due to the electric field is added, which is destabilizing in nature. Hence, the total excess free energy of PDMS films can be written as,
\begin{equation}
\label{eq:1}
\Delta G(h, V)=\Delta G_\mathrm{LW}(h) +\Delta G_\mathrm{AB}(h) +\Delta G_\mathrm{EL}(h, V) + \frac{c_\mathrm{OTS}}{h^8}
\end{equation}
where $\Delta G_\mathrm{LW}(h)$ is the Lifshitz-van der Waals interaction, $\Delta G_\mathrm{AB}(h)$ is the acid-base interaction, and $\Delta G_\mathrm{EL}(h, V)$ is the electric contribution. The last term is due to the short-range Born repulsion, $c_{\mathrm{OTS}}$ is the strength of the short-range interaction (1.8$\times$10$^{-77}$ Jm$^6$), and $h$ is the thickness of as prepared PDMS films (before depositing aqueous drops).\cite{seemann2001gaining} Upon applying an ac electric field, the Lifshitz-van der Waals and acid-base contributions remain independent of the applied voltage and the electrical contribution depends on the applied voltage as V$^2$. \cite{verma2005electric, bhatt2023electric} Furthermore, to investigate the role of frequency on dewetting dynamics, only the electrical contribution is taken into account since the Lifshitz-van der Waals and acid-base contributions to the the excess free energy is much smaller. 

\begin{figure*}[htbp]
	\centering
		\includegraphics[width=0.9\textwidth]{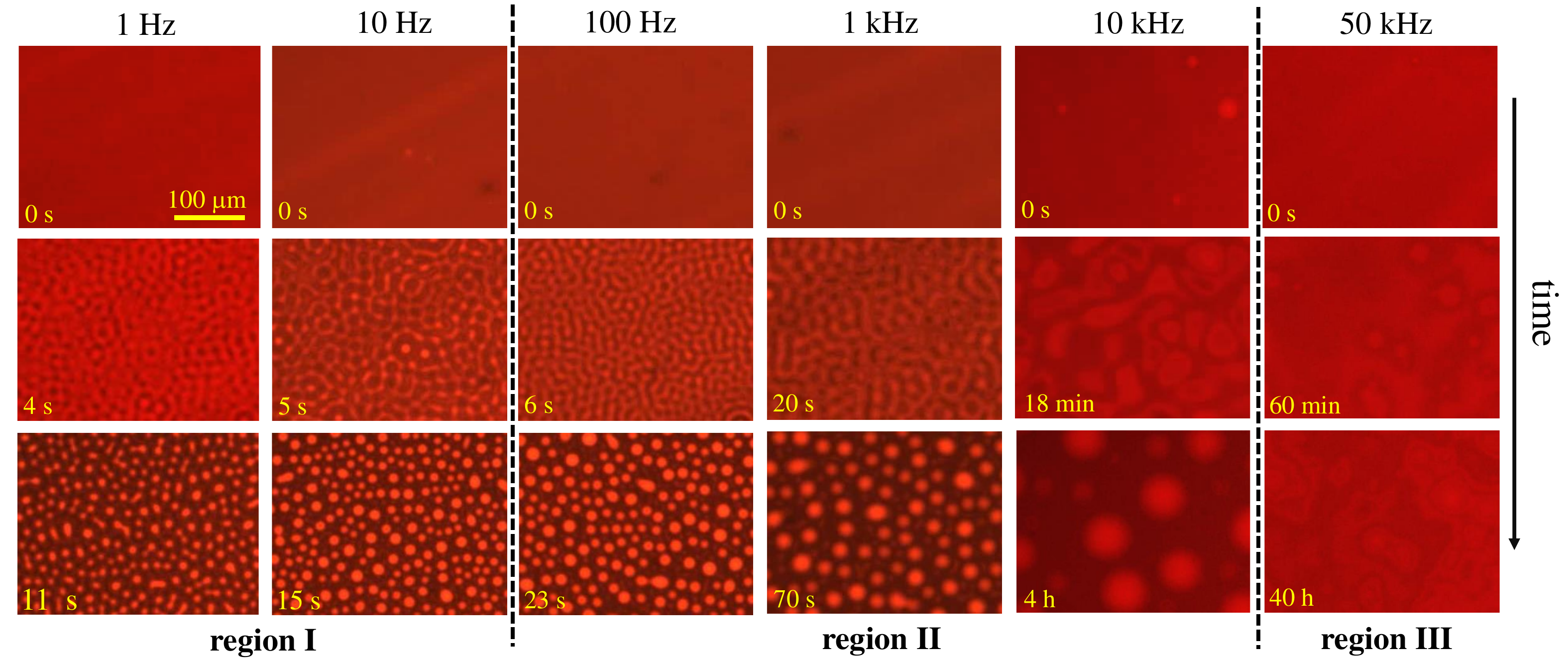}
	\caption{Fluorescent optical micrographs showing the dewetting dynamics of thin PDMS films at different frequency of the applied ac voltage. First row corresponds to the initial films before applying the voltage, and the second and third rows represent intermediate growth stage of surface capillary waves and the final dewetted morphology, respectively. Scale bar for all the micrographs is 100 $\upmu$m.}
	\label{fig:Figure2}
\end{figure*}

During experiments, fixed ac voltages of 20 V and 30 V with varying frequencies from 0.1 Hz to 50 kHz are applied across thin PDMS film. Before applying the voltage, only thermal fluctuations are present at the film-drop interface, and the film is stabilized due to the long-range Lifshitz-van der Waals interaction. After applying the voltage, surface capillary waves with wavelengths larger than the critical one appear, which tend to destabilize the thin PDMS films.\cite{bhatt2023electric} As a result, the uniform PDMS films finally dewet into multiple smaller dewetted droplets. Figure \ref{Fig:Figure 1}(b) schematically shows different dewetting morphologies for different frequencies of the applied ac electric field with corresponding fluorescent optical micrographs in Figure \ref{fig:Figure2}. We observe that at low frequencies ($f<$ 10 Hz) the large aqueous drop and the dewetted droplets do not stay static but oscillate or shift, whereas at intermediate frequencies (10 Hz $<f<$ 10 kHz) stable dewetting pattern is observed. At high frequencies ($f>$ 10 kHz), the PDMS films remain stable and no dewetting is observed. Therefore, based on the behavior of dewetted droplets as a function of applied ac frequencies, the entire dewetting process can be divided into three regions as discussed here. 

\begin{figure}[htbp!]
	\centering
		\includegraphics[width=0.4\textwidth]{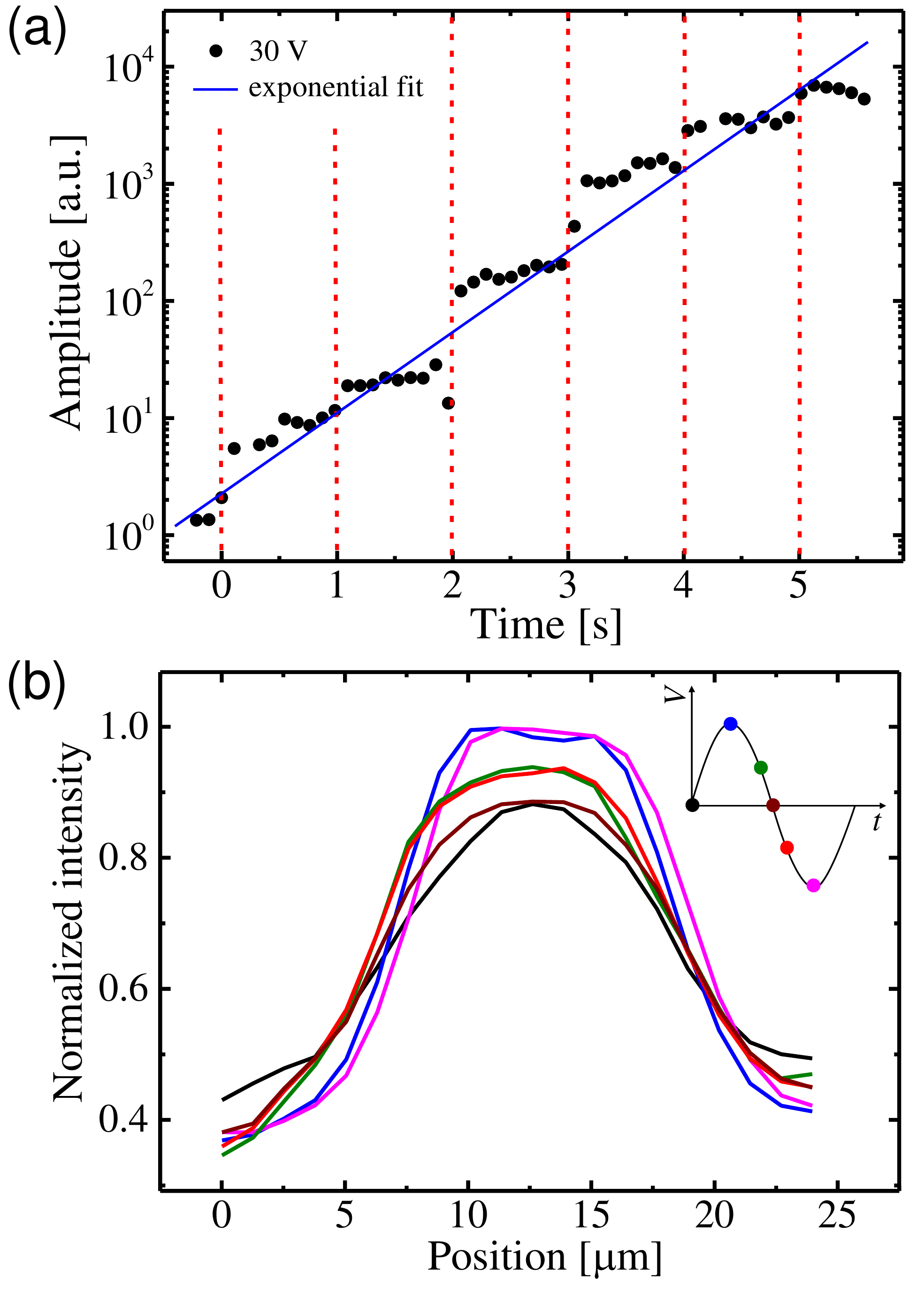}
	\caption{(a) Graph showing the amplitude of the fastest growing mode ($k_{m}$) for 1 Hz 30 V ac voltage with time. It is clear that the amplitude does not grow continuously rather grows in steps of 1 s interval. (b) Normalized intensity of a dewetted drop at 1 Hz 30 V ac voltage showing the variation in the drop height during different times of the applied ac signal (shown in inset).}
	\label{fig:Figure3}
\end{figure}

Figure \ref{fig:Figure2} shows fluorescent optical images of dewetting dynamics at different frequencies of applied ac voltage of 30 V at initial (top row), intermediate (middle row) and final (bottom row) times. The intensity variation of the fluorescent images were used to analyze the amplitude and growth of surface capillary waves during dewetting. It is clear that even for a fixed applied voltage, the final dewetting pattern depend upon the frequency of the applied voltage and this is the main objective of the study in this work. For applied ac voltage with frequencies below 10 Hz, top aqueous drops, thin PDMS films and final dewetted droplets, they all oscillate as they can respond to the slowly varying low frequency electric field. 100 $\upmu$m$\times$100 $\upmu$m area of the fluorescent optical micrographs shown in Figure \ref{fig:Figure2} were used to analyze the dewetting dynamics. During instability growth at 1 Hz frequency, the amplitude of surface capillary waves (derived from the power spectral density) are shown in the semi-logarithmic plot in Fig. \ref{fig:Figure3} (a). \cite{bhatt2022dewetting} It is clear that the amplitude does not grow continuously but rater grows in steps of 1 s interval. 
However, it was observed that during 1 s period, the amplitude only increases or remains constant but does not decreases since rewetting process is relatively slower than dewetting. \cite{edwards2016not, edwards2020viscous, bhatt2022dewetting} As a result, the amplitude overall increases or remains constant during the dewetting growth. Upon complete dewetting, the dewetted droplets also oscillate at the same rate. It has been already shown that for spinodal type dewetting, the amplitude follows exponential growth ($A(t)\propto\mathrm{exp}(t/\uptau_\mathrm{i})$) as also shown in Fig. \ref{fig:Figure3}(a). Here $\uptau_\mathrm{i}$ represents the instability growth time constant. From the fitting, the time constant of dewetting for 1 Hz frequency is $\uptau_\mathrm{i} = 1.4 \pm 0.2$ s. Subsequently, when the dewetting is completed, i.e., the capillary waves break up into small dewetted droplets and follow the nearest neighbor distribution same as the wavelength of the capillary waves, the formed dewetted droplets also show oscillatory behavior. Figure \ref{fig:Figure3} (b) shows the actual amplitude (in the form of normalized fluorescent intensity) of the oscillation of a dewetted droplet at different time corresponding to the input ac voltage (shown in inset). Since the excess free energy, responsible for the dewetting, depend on V$^2$, capillary waves reach maxima and minima twice in one voltage cycle. As a result, the capillary waves during dewetting also appear to oscillate at 1 Hz rate. For the low frequencies, the top aqueous drops also show oscillations, which is reflected as the change in the contact angle of the drops. The apparent contact angle of the aqueous drops for 1 Hz frequency at 30 V, fluctuates between $\sim$40$^{\circ}$ to $\sim$60$^{\circ}$ with the interval of 1 s. 

In region II with frequency from 10 Hz to 10 kHz, stable dewetting pattern are observed as shown in Fig. \ref{fig:Figure2}. It is also clear from the fluorescent images that the wavelength of the capillary waves in this region increases with the applied frequency. It was observed that 10 Hz frequency is the boundary between the region II and III. Around this frequency, the final dewetted drops oscillates and also move leading to the final dewetting pattern where the separation and size of dewetted drops also change due to the motion and coalescence of the dewetted droplets. Above 10 Hz frequency, this effect is negligible since the dewetted droplets cannot respond anymore. It should also be noted that the time taken in complete dewetting increases with increasing frequency of the applied ac voltage. In high frequency region with $f>$ 10 kHz, either dewetting takes very long time or the films do not dewet at all. For 10 kHz frequency, it takes four hours in complete dewet, however at 50 kHz, the PDMS films do not dewet and only small fluctuations are present at the drop-film interface.

\begin{figure}[htbp!]
	\centering
		\includegraphics[width=0.4\textwidth]{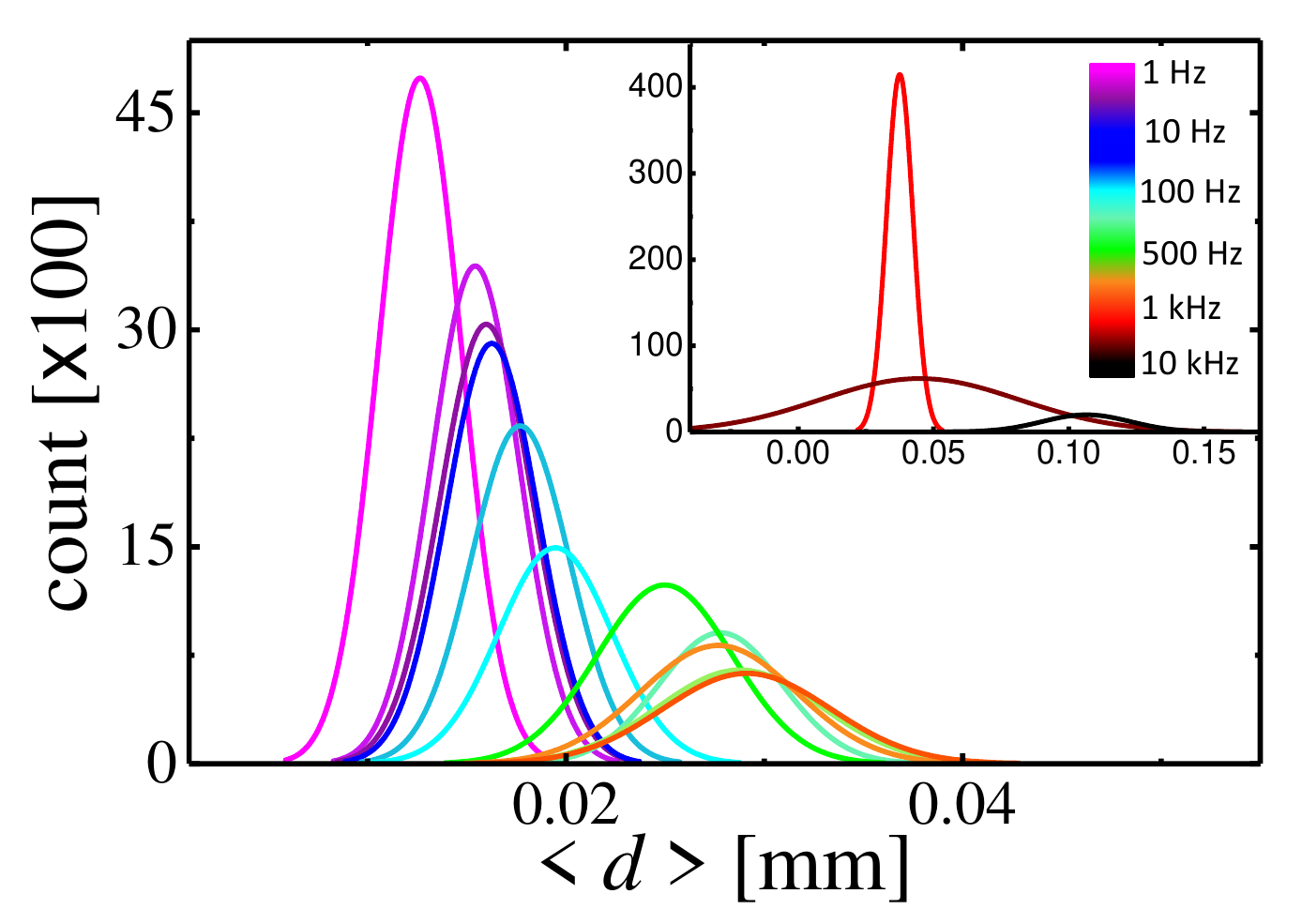}
	\caption{Histogram of dewetted droplets showing the nearest neighbor distance ($<d>$) between dewetted droplets with different applied ac frequencies at 30 V. Inset shows the histogram for higher frequencies.}
	\label{fig:Figure4}
\end{figure}

The nearest neighbor distance histogram of the final dewetted droplets for applied frequency range from 1 Hz to 10 kHz is shown in Figure \ref{fig:Figure4} for 30 V. The figure inset shows histogram for higher frequencies. It is clear from the figure that as the frequency of the applied ac voltage increases, total number of dewetted droplets decreases, however the separation between the droplets increases. Experiments were also performed at very low frequency of 0.1 Hz but the dewetted droplets oscillate and coalesce with each other so quickly that the histogram cannot be plotted for that frequency.
The role of the frequency of the applied ac voltage on dewetting dynamics and the final morphology of dewetted droplets can be explained in terms of the change in the transient dielectric properties of thin PDMS films. When a dielectric material is exposed to an external ac electric field, the dielectric molecules get polarized via different mechanism such as dipolar, atomic, ionic, electronic, and space charge polarization. Such polarizations of dielectric molecules depend on the applied frequency; at very low frequencies, all types of polarization are responsible for the change in the dielectric properties, however, at higher frequencies only a few remains left. In our experiments, we are limited to frequencies less than equal to 10 kHz, so it is possible that the polarization of the dielectric films is due to all such contributions. 

\subsection{Dielectric relaxation of PDMS molecules}
\begin{figure*}[htbp!]
	\centering
		\includegraphics[width=1.00\textwidth]{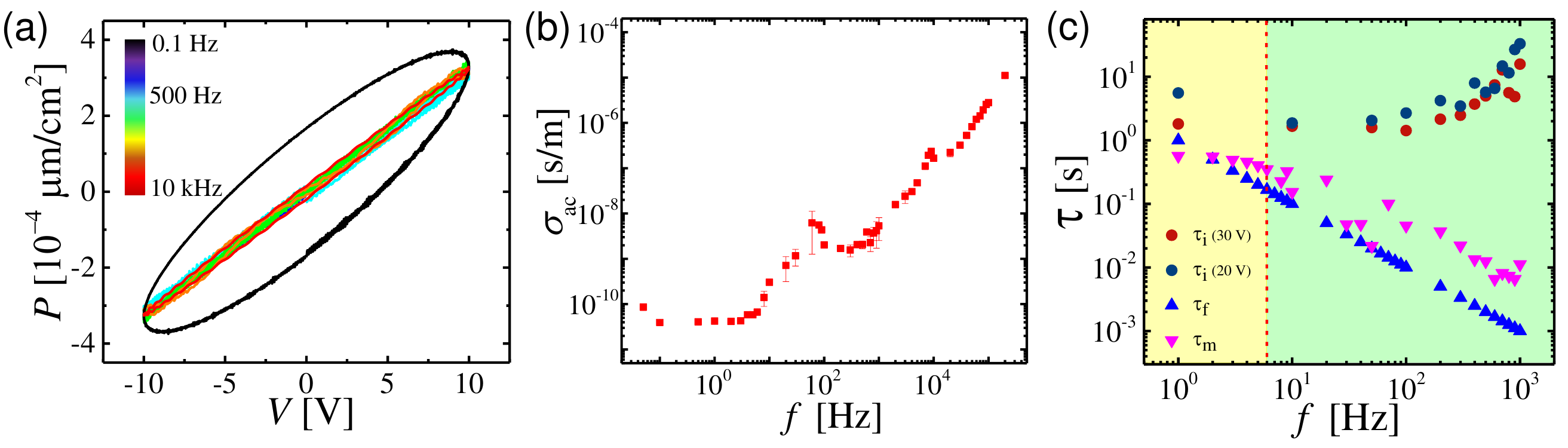}
	\caption{Dielectric response of PDMS in ac electric field; (a) polarization $P$ versus voltage $V$ at 0.1 Hz to 10 kHz applied frequency, (b) ac conductivity ($\sigma$) with applied ac frequency ($f$), and (c) variation of the three-time constants $\uptau_{\mathrm{i}},\uptau_{\mathrm{m}}$, and $\uptau_{f}$, corresponding to the the instability growth, the molecular relaxation time, and applied frequency, respectively, with frequency ($f$).}
	\label{fig:Figure5}
\end{figure*}

Dielectric response of PDMS fluid was measured using a custom built set-up electrochemical cell by measuring the polarization response of PDMS molecules. When ac voltage of 20 V$_{\mathrm{PP}}$ and frequencies from 0.1 Hz to 10 kHz are applied in the cell, the polarization curve of PDMS for positive and negative half cycles are shown in Fig. \ref{fig:Figure5}(a). From the polarization curve, it is clear that PDMS is a linear dielectric, and the slope of the polarization curve gives the value of the dielectric constant. Dielectric relaxation of PDMS molecules with the ac applied field is responsible to the frequency dependent dewetting dynamics of thin PDMS films. In presence of ac electric field, the system contains three different time constants, namely; molecular relaxation ($\uptau_\mathrm{m}$), instability growth ($\uptau_\mathrm{i}$), and applied frequency ($\uptau_\mathrm{f}$). \cite{gambhire2012role} These time constants can be changed by tuning PDMS properties such as viscosity, surface tension, and conductivity and the frequency of the applied ac voltage. The molecular relaxation time constant depends on the ac conductivity of the PDMS as $\uptau_\mathrm{c} = \epsilon\epsilon_{\mathrm{0}}/\sigma_{\mathrm{ac}}$ and PDMS being a dielectric in nature, its total conductivity can be written $\sigma^{\star}=\sigma_{\mathrm{dc}}+ \sigma_{\mathrm{ac}}$, where $\sigma_{\mathrm{dc}}$ and $\sigma_{\mathrm{ac}}$ are the dc and ac conductivities, respectively. \cite{raju2003dielectrics} The ac conductivity of a dielectric is further defined as $\sigma_{\mathrm{ac}}=2\pi\epsilon_{\mathrm{0}}\epsilon^{\mathrm{\prime\prime}}f$, where $\epsilon^{\mathrm{\prime\prime}}$ is the imaginary part of the complex dielectric constant. 
The ac conductivity ($\sigma_{\mathrm{ac}}$) is calculated experimentally from the polarization ($P$) versus applied voltage ($V$) curve and is shown in Figure \ref{fig:Figure5}(b). It is clear that $\sigma_{\mathrm{ac}}$ increases by five orders of magnitude with frequency variation from 0.1 Hz to 10 kHz. This clearly shows that the PDMS becomes more conductive with increasing frequency of ac fields. As a result, dewetting of thin PDMS films becomes weaker showing larger wavelengths at higher frequencies. 

To understand the interplay of various time constants, they were calculated from different experimental observations. The molecular relaxation time constant can be calculated by substituting the ac conductivity in the relation $\uptau_\mathrm{m} = \epsilon\epsilon_{\mathrm{0}}/\sigma_{\mathrm{ac}}$. The instability growth time constant is calculated from the slope of the growth of the amplitude of the capillary wave, $A(t)\propto\mathrm{exp}(t/\uptau_\mathrm{i})$ (cf. Fig. \ref{fig:Figure3}(a)), and the applied frequency time constant is simply the inverse of the ac frequency, i.e. $\uptau_\mathrm{f} = 1/f$. All the three time constants are plotted in Figure \ref{fig:Figure5}(c) and can be divided into two regions. For frequency below 10 Hz, all the time constants nearly the same. Hence, the top aqueous drop and the dewetted droplets oscillate at the same rate. For frequencies greater than 10 Hz, $\uptau_\mathrm{m}$ is always greater than $\uptau_\mathrm{f}$, hence PDMS acts as an ideal dielectric resulting in homogeneous and stable dewetting. At even higher frequencies ($f>$10 kHz), there occurs a crossover between $\uptau_\mathrm{m}$ and $\uptau_\mathrm{f}$ resulting into $\uptau_\mathrm{m}$ smaller than $\uptau_\mathrm{f}$. As a result, PDMS starts behaving like a leaky dielectric or conductor and thin PDMS films do not dewet anymore. \cite{gambhire2012role}

\subsection{Modified linear stability analysis}
Due to the electrical contribution to dewetting, conventional linear stability analysis due to only short and long-range forces needs to be modified to include the frequency dependent electrical field induced dewetting. Force due to the external electric field dominates the surface tension force and destabilizes thin PDMS films. Theoretically, the surface evolution and the wavelength of the fastest-growing mode are derived from the continuity and Navier-Stokes equations. In the Navier-Stokes equation, the body forces are due to the van der Waals, acid-base, and external electric field. The Reynolds and Capillary numbers are assumed to be very small, so all the inertial effects can be safely ignored. The electric field-induced instability is long-wave type, and the characteristic length $\epsilon=h_{\mathrm{0}}/L\rightarrow0$, where, $h_{\mathrm{0}}$ is initial film thickness and $L$ is the characteristic lateral length. Using long-wave approximation for Newtonian fluid with no-slip boundary condition, the Navier-Stokes equation can be simplified to a thin film equation as, \cite{verma2005electric}
\begin{equation}
\label{eq:2}
3 \eta \left(\frac{\partial h(x,t)}{\partial t}\right) -\nabla \cdot[h^{3}\nabla P]=0
\end{equation}
where $h(x,t)$ and $\eta$ are the thickness and viscosity of PDMS films, respectively, and $P$ is the total pressure across the PDMS films. In the total pressure, all the contributions of body forces, such as pressure due to the curvature, intermolecular conjoining pressure, and Maxwell stress tensor due to electric field, are included. By taking ansatz $h = h_{\mathrm{0}}+\epsilon\mathrm{exp}(\omega t-ikx)$, where $\omega$ is the growth rate and $k$ is the wavenumber of the fastest growing capillary waves and Eq. \ref{eq:2} can be linearized as 
\begin{equation}
\label{eq:3}
\omega = -\frac{k^{2}h_\mathrm{0}^{3}}{3 \eta }\left(k^{2}\gamma_\mathrm{PD}+\Delta G^{\prime\prime}(h=h_{\mathrm{0}})\right)
\end{equation}
where $\gamma_\mathrm{PD}$ is the interfacial tension between PDMS and aqueous drop. The wavelength corresponding to the fastest growing mode is calculated by taking $\mathrm{d}\omega/\mathrm{d}k=0$, as
\begin{equation}
\label{eq:4}
\lambda_\mathrm{m} = 2 \pi \left(-\frac{\Delta G^{\prime\prime}({h=h_\mathrm{0}})}{2\gamma_\mathrm{PD}}\right)^{-1/2}
\end{equation}
The total excess free energy for the present three-layer system can be written as 
\begin{equation}
\begin{split}
\label{eq:5}
\Delta G(h, V)\,=+\frac{A_{\mathrm{OTS/PDMS/drop}}-{A_{\mathrm{SiO_2/PDMS/drop}}}}{12 \pi(\,{h+d_{\mathrm{OTS}}})^{2}}\\
+\frac{A_{\mathrm{SiO_2/PDMS/drop}}-A_{\mathrm{Si/PDMS/drop}}}{12 \pi (\,{h+d_{\mathrm{OTS}}+d_{\mathrm{SiO_2}}})^{2}}\\
-\frac{A_{\mathrm{OTS/PDMS/drop}}}{12 \pi h^{2}}
+S_{\mathrm{P}}\,\mathrm{exp}(\,\frac{{d_{\mathrm{min}}}-h}{l})\\
-\frac{1}{2}\frac{\epsilon_\mathrm{0}\epsilon_\mathrm{SiO_2}}{d_\mathrm{SiO_2}}\frac{1}{(1+\frac{d_\mathrm{OTS}\epsilon_\mathrm{SiO_2}}{d_\mathrm{SiO_2}\epsilon_\mathrm{OTS}}+\frac{h\epsilon_\mathrm{SiO_2}}{d_\mathrm{SiO_2}\epsilon_\mathrm{PDMS}})}V^2 
+ \frac{c_\mathrm{OTS}}{h^8}
\end{split}
\end{equation}
In Eq. \ref{eq:4}, $\Delta G^{\prime\prime}({h=h_\mathrm{0}})$ can be calculated using Eq. \ref{eq:1}. The total excess free energy includes $h$ dependency for the short and long range interactions and $V$ dependency for the applied electric field. However, the dependency of frequency of the applied ac voltage is not captured in the linear stability analysis. In Eq. \ref{eq:5}, $\Delta G$ is a function of $h$, $V$, and $f$. The first three terms in Eq. \ref{eq:5} are due to the van der Waals interaction between the three dielectric layers PDMS, OTS, and SiO$_2$, and depends on the Hamaker constant, which is a function of frequency and can be written as \cite{israelachvili2011intermolecular}
\begin{equation}
\label{eq:6}
\begin{split}
A=\frac{3}{4}k_\mathrm{B}\mathrm{T} \frac{(\varepsilon_\mathrm{L}-\varepsilon_\mathrm{D})}{(\varepsilon_\mathrm{L}+\varepsilon_\mathrm{D})}\frac{(\varepsilon_\mathrm{S}-\varepsilon_\mathrm{L})}{(\varepsilon_\mathrm{S}+\varepsilon_\mathrm{L})}\\
+\frac{3\pi \hbar \nu_\mathrm{e}}{4 \sqrt{2}} \frac{(n_\mathrm{L}^2-n_\mathrm{D}^2 )(n_\mathrm{S}^2-n_\mathrm{L}^2 )}{(\sqrt{(n_\mathrm{D}^2+n_\mathrm{L}^2 )(n_\mathrm{S}^2+n_\mathrm{L}^2 )} [\sqrt{n_\mathrm{L}^2+n_\mathrm{D}^2}+\sqrt{n_\mathrm{S}^2+n_\mathrm{L}^2}])}
\end{split}
\end{equation}
where subscripts L, D, and S represent PDMS film, aqueous drop, and substrate phase, respectively, and $\varepsilon$, $n$ are the dielectric constant and refractive index of the materials. The first term on the right hand side of Eq. \ref{eq:6} is the dc term for zero frequency, and the second term is due to the non-zero frequency contribution. The acid-base contribution to the excess free energy is negligibly small, hence can be ignored. 

\begin{figure}[htbp!]
	\centering
		\includegraphics[width=0.45\textwidth]{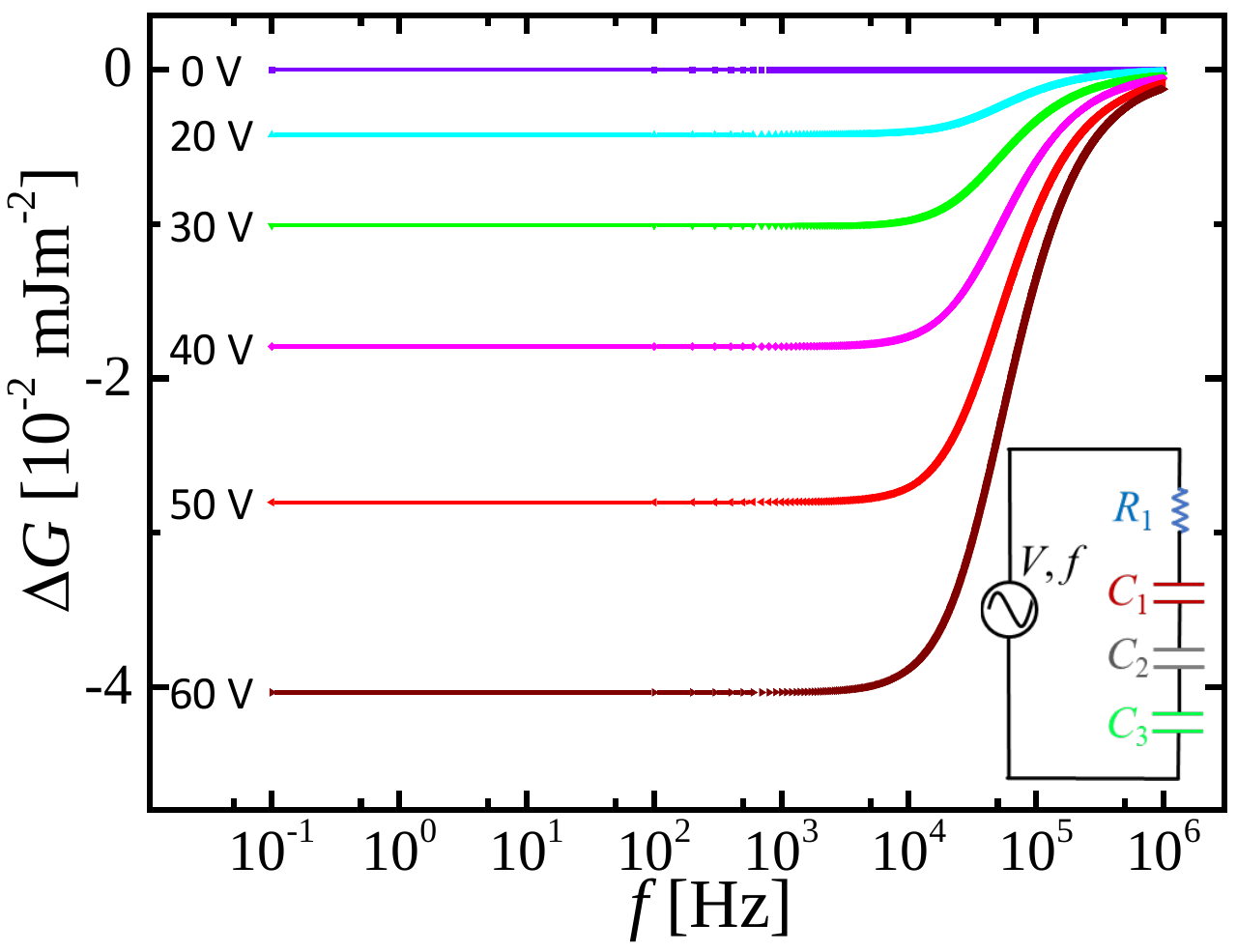}
	\caption{Modified total excess free energy ($\Delta G$) of the PDMS film underneath the aqueous drop after including the frequency effect on the interactions. Inset shows the equivalent ac electric circuit of the experimental system.}
	\label{fig:Figure6}
\end{figure}

Assuming that the dielectric layers of PDMS, OTS and SiO$_2$ form parallel plate capacitors and top aqueous drop offers a series resistance, the equivalent electric circuit will be their series combination, as shown in the inset of Fig. \ref{fig:Figure6}. 
The electric contribution to the total excess free energy can be written as,  
\begin{equation}
\label{eq:7}
\Delta G_\mathrm{EL}(h, V, f) = -\frac{1}{2}C_{\mathrm{total}}V^2_{rms} 
\end{equation}
If the capacitance of PDMS, OTS and SiO$_2$ layers are represented by C$_{1}$, C$_{2}$, and C$_{3}$, respectively, then 
\begin{equation}
\label{eq:8}
\frac{1}{C_{\mathrm{total}}} = \left(\frac{1}{C_{1}}+\frac{1}{C_{2}}+\frac{1}{C_{3}}\right)
\end{equation}
All the three capacitors charge and discharge via the series resistor $R_{\mathrm{1}}$ with respective time constants $\uptau_{\mathrm{PDMS}}=R_{\mathrm{1}} C_{\mathrm{1}}$, $\uptau_{\mathrm{OTS}}=R_{\mathrm{1}} C_{\mathrm{2}}$, and $\uptau_{\mathrm{SiO_{2}}}=R_{\mathrm{1}} C_{\mathrm{3}}$. So the Eq. \ref{eq:7} can be rewritten as 
\begin{equation}
\label{eq:9}
\Delta G_\mathrm{EL}(h, V, f) = -\frac{1}{4\pi f} \frac{1}{\left|\mathrm{Z_{total}}(h, V,\omega)\right|}V^{2}
\end{equation}
where, $\mathrm{Z_{total}}(h, V, f)$ is the total impedance of the equivalent circuit and can be calculated as $\mathrm{Z_{total}}(h, V,f) = \mathrm{Z_{C_{1}}}+\mathrm{Z_{C_{2}}} +\mathrm{Z_{C_{3}}}+\mathrm{Z_{R_{1}}}$. The squared value of total impedance is 
\begin{equation}
\label{eq:10}
\begin{split}
{\left|\mathrm{Z_{total}}(h,V,f)\right|}^2 = \frac{1}{4 {\pi}^2 f^2 \varepsilon_{\mathrm{0}}^2}\left(\frac{h^2}{\varepsilon^{\star 2}_{\mathrm{PDMS}}}+\frac{d_{\mathrm{OTS}}^2}{\varepsilon^{\star 2}_{\mathrm{OTS}}}+\frac{d_{\mathrm{SiO_2}}^2}{\varepsilon^{\star 2}_{\mathrm{SiO_2}}}\right)\\
+\frac{1}{\varepsilon_{\mathrm{0}}^2}\left(\frac{\uptau_{\mathrm{PDMS}}h}{\varepsilon^{\star}_{\mathrm{PDMS}}}+\frac{\uptau_{\mathrm{OTS}}d_{\mathrm{OTS}}}{\varepsilon^{\star}_{\mathrm{OTS}}}+\frac{\uptau_{\mathrm{SiO_2}}d_{\mathrm{SiO_2}}}{\varepsilon^{\star}_{\mathrm{SiO_2}}}\right)^2
\end{split}
\end{equation}
where $\uptau_{\mathrm{PDMS}},\uptau_{\mathrm{OTS}},\uptau_{\mathrm{\mathrm{SiO_2}}}$ are the relaxation time for PDMS, OTS and SiO$_2$, respectively, and $\varepsilon^{\star}$ correspond to the complex dielectric constant for different materials. By substituting the value of $\mathrm{Z_{total}}(h, V, f)$ in Eq. \ref{eq:9}, $\Delta G_\mathrm{EL}$ can be calculated. The relative strength of the electrical contribution to the total excess free energy is much much larger than all other contribution, hence they can be neglected and $\Delta G \approx \Delta G_\mathrm{EL}$. From Eq. \ref{eq:9}, $\Delta G$ is calculated for different applied ac frequencies and is plotted in the logarithmic scale in Fig. \ref{fig:Figure6}. The higher the applied frequency magnitude of the free energy is lower compared to the lower frequency shows the slow dewetting at a higher frequency. Similarly, the higher the applied voltage faster the dewetting dynamics.
\begin{figure}[htbp!]
	\centering
		\includegraphics[width=0.45\textwidth]{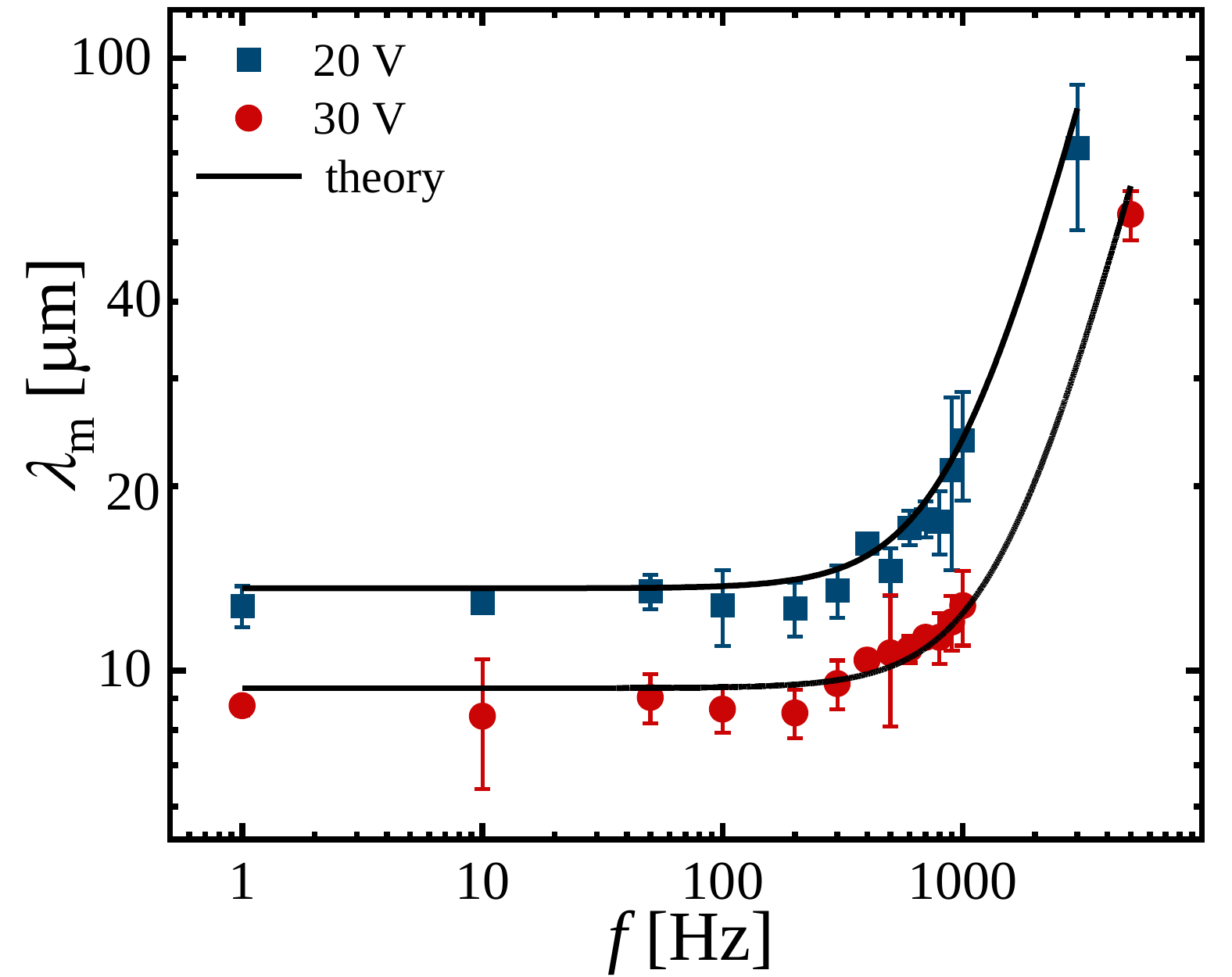}
	\caption{Wavelength of the fastest growing mode ($\lambda_{\mathrm{m}}$) with frequency ($f$) during dewetting of thin PDMS films at 20 V and 30 V. The solid lines represent the modified linear stability analysis (obtained from Eq. \ref{eq:9}) fitted to the experimental data.}
	\label{fig:Figure7}
\end{figure}
By substituting the value of $\Delta G^{\prime\prime}(h, V, f)$ into Equation \ref{eq:4}, the wavelength calculated from linear stability analysis can be modified. The modified linear stability analysis is fitted to the experimental data shown in Figure \ref{fig:Figure7}, and it fits very well with the experimental data with the R$^2=0.9$ for both the 20 V and 30 V applied voltage. During fitting $\lambda_m$ from Eq. \ref{eq:9} and Eq. \ref{eq:10}, the molecular relaxation time constant of PDMS, $\uptau_{\mathrm{PDMS}}$ was used as a fitting parameter which agreed very well from the value obtained using the polarization experiment (cf. Figure \ref{fig:Figure5}) and also using the theoretical relation $\uptau= ((\varepsilon_{\mathrm{s}}+2)/(n^2+2))\uptau_{\mathrm{c}}$. Here $\uptau{_\mathrm{c}}= (4\pi\eta a^3)/(kT)$ and $k$, T, $\varepsilon_{\mathrm{s}}$, $n$ are Boltzmann constant, absolute temperature, dc dielectric constant and refractive index of the material, respectively. \cite{raju2003dielectrics} 

These observation and analysis confirm that the modified linear stability analysis can be used for any applied voltage and frequency to predict the dewetting dynamics and the wavelength of the fastest-growing mode. 

\section{Conclusion}
In summary, we found that the wavelength of the fastest-growing mode during dewetting PDMS film underneath an aqueous drop depends on the frequency of the applied external electric field. The linear stability theory explains the effect of the strength of the external field on the fastest-growing wavelength, but it fails to explain the frequency effect on dewetting dynamics. In this theory, the dc voltage is replaced by the root mean square voltage for ac applied field. For an ac field, three-time scales are involved during dewetting, and the interplay between these leads to different parameters of dewetting. We have experimentally observed that with the increase in the frequency of the applied field, the wavelength of the fastest-growing mode of the capillary waves increases. For voltages $<$ 10 Hz, dewetting shows the frequency response during dewetting, and it also oscillates after droplet formation. But dewetted droplets are stable after formation for higher frequencies $>$ 10 Hz. For very high frequencies $>$ 10 kHz, there is no dewetting or very large wavelength, which is very large from the experimental length scale. Also, it takes a very large time compared to the experimental time scale to dewet. All the time scales are almost equal for very low frequency is responsible for the oscillatory behavior at a lower voltage. No dewetting at a higher frequency is because the PDMS becomes conductive at higher frequency while aqueous liquid becomes dielectric in nature. Similarly, the excess free energy also depends on the frequency, which also changes with the frequency of the applied field. In the linear stability analysis, the contribution due to applied frequency is not included; it only predicts the fastest growing wavelength for higher frequency where voltage is similar to root mean square voltage.

We developed a modified linear stability analysis by adding the time constants associated with the different dielectric layers and the complex dielectric constant for different materials. The modified theory predicts the fastest-growing wavelength for all the frequency ranges from 1 Hz to 10 kHz. And it also shows that the wavelength diverges at very high frequencies $>$ 10 kHz, which is also observed in our experiments. From this frequency-dependent dewetting, one can also calculate the relaxation time the lubricating fluid polymer with the external applied electric field.

\section{acknowledgement} 
KK acknowledges the funding support from SERB, New Delhi (Project no. CRG/2019/000915) and DST, New Delhi, through its Unit of Excellence on Soft Nanofabrication at IIT Kanpur.  

\bibliography{Manuscript}

\end{document}